\documentstyle[epsfig,equation]{aipproc2}
\begin{document}

\def \simlt{\lower.5ex\hbox{$\; \buildrel < \over \sim \;$}}
\def \simgt{\lower.5ex\hbox{$\; \buildrel > \over \sim \;$}}
\def\lsim{\lower2pt\hbox{$\buildrel{<}\over{\sim}$}}
\def\gsim{\lower2pt\hbox{$\buildrel{>}\over{\sim}$}}
\def\rr{{\rm r}}
\def\cc{{\rm c}}
\def\mm{{\rm m}}
\def\ss{{\rm s}}
\newcommand{\FF}{{\cal F}}
\newcommand{\cd}{\cdot}
\newcommand{\ip}{\int_0^{2\pi}}
\newcommand{\al}{\alpha}
\renewcommand{\b}{\beta}
\newcommand{\de}{\delta}
\newcommand{\De}{\Delta}
\newcommand{\ep}{\epsilon}
\newcommand{\ga}{\gamma}
\newcommand{\Ga}{\Gamma}
\newcommand{\ka}{\kappa}
\newcommand{\io}{\iota}
\newcommand{\La}{\Lambda}
\newcommand{\la}{\lambda}
\newcommand{\Om}{\Omega}
\newcommand{\om}{\omega}
\newcommand{\si}{\sigma}
\newcommand{\Si}{\Sigma}
\newcommand{\th}{\theta}
\newcommand{\vth}{\vartheta}
\newcommand{\vph}{\varphi}
\newcommand{\ra}{\rightarrow}
\newcommand{\tr}{\mbox{tr}}
\newcommand{\hor}{\mbox{hor}}
\renewcommand{\baselinestretch}{1.01}
\newcommand{\bea}{\begin{eqnarray}}
\newcommand{\eea}{\end{eqnarray}}
\newcommand{\dd}{\partial}
\def\laq{\raise 0.4ex\hbox{$<$}\kern -0.8em\lower 0.62
ex\hbox{$\sim$}}
\def\double{\baselineskip 24pt \lineskip 10pt}
\newcommand{\sx}{\sigma}
\newcommand{\sei}{\sigma_8}
\newcommand{\sxa}{\sigma_1}
\newcommand{\sxb}{\sigma_2}
\newcommand{\pha}{\phi_1}
\newcommand{\phb}{\phi_2}
\newcommand{\Pha}{\Phi_1}
\newcommand{\Phb}{\Phi_2}
\newcommand{\Phib}{\bar{\Phi}}
\newcommand{\Phab}{\bar{\Phi}_1}
\newcommand{\Phbb}{\bar{\Phi}_2}
\newcommand{\mpl}{m_{Pl}}
\newcommand{\Mpl}{M_{Pl}}
\newcommand{\lx}{\lambda}
\newcommand{\Lx}{\Lambda}
\newcommand{\ex}{\epsilon}
\newcommand{\be}{\begin{equation}}
\newcommand{\ee}{\end{equation}}
\newcommand{\een}{\end{subequations}}
\newcommand{\ben}{\begin{subequations}}
\newcommand{\beq}{\begin{eqalignno}}
\newcommand{\eeq}{\end{eqalignno}}
\def \lta {\mathrel{\vcenter
     {\hbox{$<$}\nointerlineskip\hbox{$\sim$}}}}
\def \gta {\mathrel{\vcenter
     {\hbox{$>$}\nointerlineskip\hbox{$\sim$}}}}

\title{BOOMERanG: a scanning telescope \\ 
for 10 arcminutes resolution CMB maps} 

\author{S. Masi$^1$, P.A.R. Ade$^2$, R. Artusa$^3$, J.J. Bock$^3$, A. Boscaleri$^4$, 
B.P. Crill$^3$, P. de Bernardis$^1$, G. De Troia$^1$, P.C. Farese$^5$, 
M. Giacometti$^1$, V.V. Hristov$^3$, A. Iacoangeli$^1$, A.E. Lange$^3$, A.T. Lee$^6$, L. Martinis$^7$,
P.V. Mason$^3$, P.D. Mauskopf$^8$, F. Melchiorri$^1$, L. Miglio$^1$, T. Montroy$^5$,
C.B. Netterfield$^3$, E. Pascale$^{3,4}$, F. Piacentini$^1$, P.L. Richards$^6$, 
G. Romeo$^9$, J.E. Ruhl$^5$, 
F. Scaramuzzi$^7$}

\address{$^1$ Dipartimento di Fisica, Universita' La Sapienza, Roma, Italy \\
 $^2$ Queen Mary and Westfield College, London, UK \\
 $^3$ California Institute of Technology, Pasadena, CA, USA \\
 $^4$ IROE-CNR, Firenze, Italy \\
 $^5$ Dept. of Physics, Univ. of California, Santa Barbara, CA, USA \\
 $^6$ Dept. of Physics, Univ. of California, Berkeley, CA, USA \\
 $^7$ ENEA CRE Frascati, Italy \\
 $^8$ Dept. of Physics and Astronomy, University of Massachusets, Amherst, MA, USA \\
 $^9$ Istituto Nazionale di Geofisica, Roma, Italy }

\maketitle

\begin{abstract}
The BOOMERanG experiment is a stratospheric balloon
telescope intended to measure the Cosmic Microwave Background
anisotropy at angular scales between a few degrees and ten
arcminutes. The experiment features a wide focal plane
with 16 detectors in the frequency bands centered at
90, 150, 220, 400 GHz, with FWHM ranging between 18 and 10 arcmin.
It will be flown on a long duration (7-14 days) 
flight circumnavigating Antarctica at 
the end of 1998. The instrument was flown with
a reduced focal plane (6 detectors, 90 and 150 GHz bands,
25 to 15 arcmin FWHM) on a qualification flight
from Texas, in August 1997. 
A wide ($\sim$ 300 deg$^2$, i.e. about 5000 independent beams
at 150 GHz) 
sky area was mapped in the constellations of Capricornus, 
Aquarius, Cetus, with very low foreground contamination.
The instrument was calibrated using the CMB dipole 
and observations of Jupiter. The LDB version of the instrument has
been qualified and shipped to Antarctica.
\end{abstract}

\section*{Introduction}

The target of the BOOMERanG experiment is to make
$\ell$-space spectroscopy of the power spectrum of CMB anisotropies 
in the region of the first three acoustic peaks 
($50 \simlt \ell \simlt 800$). 
This will be obtained by multiband mapping of wide 
($\simgt$ 20000 independent pixels ) 
sky regions in the lowest foreground areas, with angular resolution 
approaching 10 arcmin FWHM and with S/N per pixel $\simgt 1$.
This target can be achieved with a scanning instrument. This class 
of instruments features a number of nice properties:

\smallskip \noindent $\bullet$ It allows simultaneous measurements 
of the CMB anisotropy spectrum over
a wide $\ell$ range, with reasonable $\ell$ resolution; 

\noindent $\bullet$ No moving parts are necessary in the optical system, thus achieving 
high reliability and avoiding offsets from chopper-synchronous signals;

\noindent $\bullet$ Such an instrument is a testbed for future satellites using
similar technologies and observation strategies, like Planck Surveyor. 

\smallskip \noindent 
Moreover, the experiment can be sub-orbital, due to recent 
progress in different areas:

\smallskip 
\noindent $\bullet$ development of very fast and sensitive bolometric 
detectors and readout electronics with very low 1/f noise;

\noindent $\bullet$ development of low-background, low-sidelobes 
microwave telescopes;

\noindent $\bullet$ Long Duration Ballooning opportunities and related payload 
technologies (CR immune hardware, stable readout, 
long duration cryogenics, stable scan-oriented attitude control system.)

\smallskip \noindent 
The quest for good angular resolution in CMB anisotropy measurements, 
with the practical need for a reasonable telescope size, 
drive us towards high frequencies, i.e. 
bolometric techniques. In fact a beam FWHM of about 10 arcmin requires 
 a telescope diameter $D \sim 3m$ in the best operation
band of HEMTs (around 40 GHz), while $D \sim 1m$ at 150 GHz, a good 
operation band for modern composite bolometers.
High frequencies require a balloon experiment to avoid
most of the atmospheric emission, thus achieving low background 
on the bolometers and low atmospheric noise. This can be obtained 
only with correct optical filtering, an 
extremely difficult and important issue of millimeter-wave
bolometric photometry.
The quest for repeated mesurements, allowing deep checks for 
systematic effects, drives us to long (few days) balloon flights. 
Long Duration Ballooning is carried out by NASA-NSBF in Antarctica.
In the case of CMB anisotropy experiments the 
advantages are as follows:

\smallskip 
\noindent $\bullet$ the long duration (7-14 days) provides 
the opportunity to check extensively for systematic effects,
the most troublesome source of errors in this type of
measurements;

\noindent $\bullet$ the long integration time per pixel 
gives high sensitivity, making CMB mapping possible
over significant sky fractions;

\noindent $\bullet$ the sun, always present, provides power supply 
through solar panels and a stable thermal environment.

\smallskip \noindent Disadvantages are also present, as listed below:

\smallskip \noindent $\bullet$ the increased cosmic rays density in
polar regions requires special, custom developed detectors;

\noindent $\bullet$ the long duration requires special cryogenic systems;

\noindent $\bullet$ the presence of the sun restricts the observations
to the antisolar regions, requires multiple sun shields to produce
a thermally stable telescope and requires very low sidelobes (-90 dB 
at 180$^o$) for the telescope;

\noindent $\bullet$ the payload is not always in the line of sight
of the telemetry base, so special communications are required, and 
the interactivity with the instrument can be somewhat reduced.

\smallskip \noindent 
We have developed and tested a payload \cite{Lange1}, \cite{Paolo1}, \cite{Silvia1}, 
which takes full advantage of the
opportunities listed above for LDB flights, and efficiently takes care
of all the related problems. The instrument is called BOOMERanG
(Balloon Observations Of Extragalactic Radiation and Geophysics).

\section*{Observation Strategy}

BOOMERanG is a scanning experiment: the beam scans the sky at constant elevation
and constant azimuth speed. Different spherical harmonic components of the 
temperature field produce different electrical frequencies in the detector. So,
in principle, $\ell$-space spectroscopy is possible using a scanning instrument and 
an averaging signal analyzer. This experimental approach has been made possible
by the development of the latest generation ultra-sensitive bolometers, achieving
a noise of $\sim 100 \mu K_{CMB} \sqrt{s}$.
In the case of scans along circles (constant elevation or declination)
there is a simple relationship between the Fourier spectrum of temperature fluctuations
measured along each circle and the usual spherical harmonic expansion coefficients on the
sphere ($C_l$) (\cite{delab}).
Given the azimuth $\phi$, the colatitude $\theta_o$ (constant along the circle),
and the measured CMB temperature $T (\theta_o,\phi)$ we define the Fourier 
transform of the CMB temperature measured along the scan
$$
\alpha_m = {1 \over 2\pi} \int_0^{2 \pi} T(\theta_o,\phi) e^{-im\phi} d\phi
$$
and the 1-D power spectrum of the temperature fluctuations 
$$
\delta_{mm'} \Gamma_m =\left< \alpha_m \alpha^*_{m'} \right>
$$
It can be shown that
$$
\Gamma_m = \sum_{l=|m|}^\infty C_l B_l^2 {\cal P}_{l,m}^2( \theta )
$$
\smallskip
\noindent This quantity is to be compared to the power 
spectra of the time series of the
signal detected by the scanning CMB telescope. 
The $m$ component of the 1-D spectrum will produce an 
electrical signal in the CMB detector at a frequency
$$
f = {{\dot \phi} \over 2\pi} m
$$
The scan speed ${\dot \phi}$ must be optimized in such a way that the 
interesting $m$ range (in our case, the region of the acoustic peaks) 
is detected in a frequency range far from 1/f noise, interferences,and
detector roll-off effects. In fig.1 we compare the system 
noise (including atmospheric effects induced by payload pendulations)
to the 1-D power spectrum of CMB anisotropies, assuming two different
scan speeds and a 12 arcmin FWHM gaussian beam. It is evident that 
with a suitable choice of the scan speed ($\simgt 2 deg/s$) the
interesting features in the CMB power spectrum can be observed
in a frequency range where detector noise is reasonably flat.

\begin{figure}[htb]
\centerline{\epsfig{file=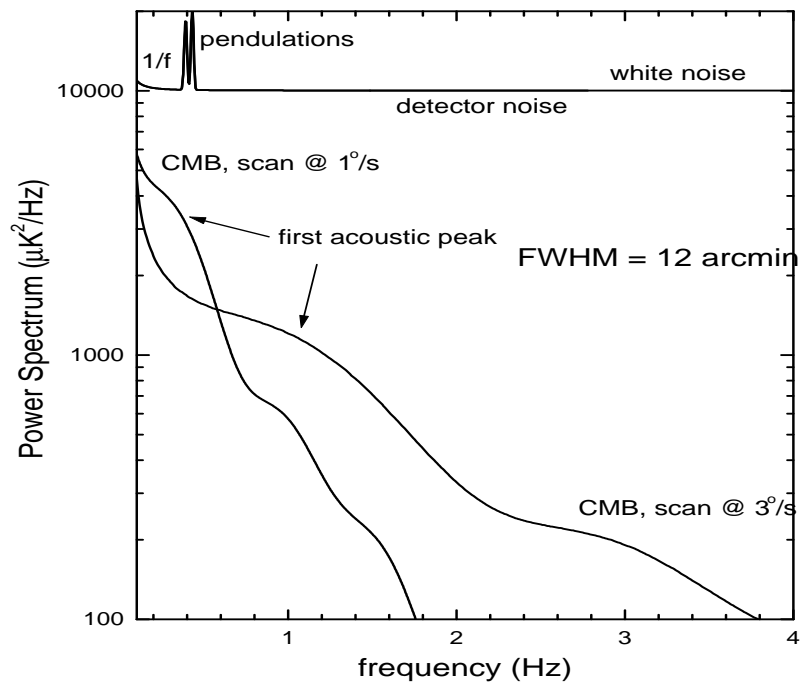,height=4.5in,width=4.5in}}
\caption{ 1-D power spectrum of CMB anisotropies (multiplied by 
the beam MTF for a 12 arcmin FWHM gaussian beam) compared to
 system noise for a scanning experiment. The goal of the 
optimization process is to move the interesting spectral
features of the signal (the "acoustic peaks" in our case)
in the frequency range where the experiment noise is flat
and low. A scan speed of $\simgt 2^o/s$ is required
to avoid 1/f noise and pendulation features.}
\label{fig1}
\end{figure}

During a day-time LDB flight the payload can observe only a 
restricted region opposite to the sun. The BOOMERanG payload
makes inversion smoothed triangular-wave azimuth scans, 
60 deg peak to peak, at constant elevation. 
This can be selected in the range $40^o \simlt e \simlt 50^o$. 

For the LDB flight the azimuth scan is centered on the Horologium 
constellation, the minimum foreground region in the 
IRAS-DIRBE maps \cite{Schl}, which happens to be close to the anti-sun 
direction during the LDB flights performed in the antarctic summer.
In this region the expected foreground fluctuation is of the order
of a few microkelvin CMB at 150 GHz.

This strategy provides us with a wide sky map (about 40 deg wide
by 30 deg high). The full map is observed
once per day for several days. The flight path is approximately
a circle at constant latitude ($\sim 80^o $ south). The
longitude drift of the experiment during the flight produces 
a variable tilt of the scan direction in the RA-dec plane, thus 
providing usefull cross linking in the dataset.

The focal plane of the experiment features two 4 x 4 arrays
of photometers (see fig.2). 
\begin{figure}[ht!]
%\centerline{\epsfig{file=fig2.eps,height=3in,width=4.5in}}
\centerline{\epsfig{file=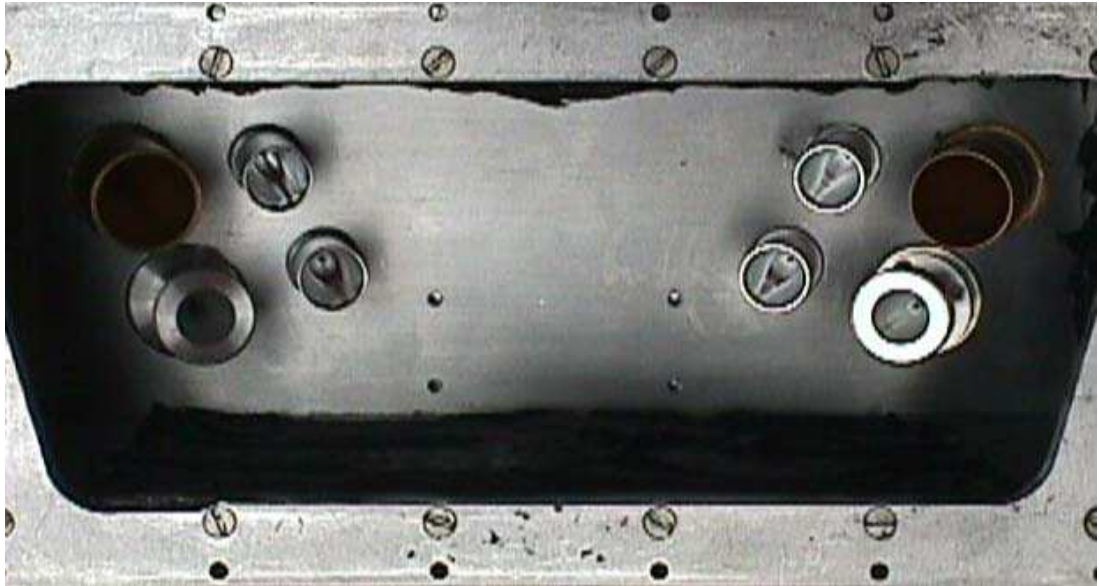}}
\caption{ Photograph of the BOOMERanG focal plane array of
receiver horns. The input diameter of the smaller horns is 6 mm.
The horns correspond to the following channels: 
top row: B90A, BXXXA1, BXXXB1, B90B;
bottom row: B150A, BXXXA2, BXXXB2, B150B. The numbers here
label the operating fequency in GHz for single mode horns; 
XXX is the input of a 3-color (150, 220, 400 GHz) photometer.}
\label{fig2}
\end{figure}

So we have three levels of sources (or structures) confirmation: 
after the first scan,
the reversed time domain structure must be evident in 
the signal from the same detector scanning back to 
the start position after few tens of seconds; 
moreover the same structure must be evident in the 
symmetric detector with the same 
frequency band active in the same row 
of the considered one (first level confirmation). 
The same structure must be evident in the signals 
from the second row of detectors, scanning the 
same sky strip a few minutes after the first row, due to the 
daily sky rotation (second level confirmation).
The same structure must be evident in the signals 
from the scans measured the subsequent day at the same time
(Third level confirmation). 
First and second level confirmations are effective 
in discerning sky signals from system instabilities 
(thermal or radiofrequency), atmospheric fluctuations and scan 
synchronous effects; third level confirmation is useful to discern 
sky signals from sidelobes artifacts (ground or sun pick-up). 

Our instrument features multiband photometers. This allows us
to analyze spectrally the detected structure, 
comparing it to the structures seen in the other spectral bands. 
This is an extremely powerful method to recognize the spectral 
signature of CMB anisotropies, which is greatly different from that
of dust and synchrotron foregrounds and from that of atmospheric 
emission.

\section*{The Instrument}

The BOOMERANG experiment is a millimeter wave
telescope, with a bolometric receiver working
in a long-duration cryostat at 0.3 K for 15 days.
The instrument is part of the BOOMERanG/MAXIMA collaboration,
sharing similar basic technologies (detectors and
ACS, see the paper from A. Lee in these proceedings). 
BOOMERanG has been optimized for the peculiar
requirements of Antarctic long duration ballooning (LDB).
A diagram of the optical system is presented in fig.3.
The primary mirror of the telescope (a 1.3 m 
diameter, 1.5 m focal length, off-axis
paraboloid) is at ambient temperature. It is
made out of 6061 aluminum, as is the entire telescope
frame. At the focus of this telescope, the multiband receiver 
features beams ranging from 12 to 20 arcmin FWHM
(depending on the channel and on the configuration).
The telescope is protected by an earth shield
and by two large sun shields, which allow operation
of the system in the range $\pm 60^o$ from the
anti-solar azimuth. Off axis radiation at 150 GHz
is undetected at the -85dB level.
The secondary and tertiary mirrors are cooled
to 2K inside the cryostat. Radiation enters the
cryostat through a 50 $\mu m$ thick polypropylene
window, and goes through blocking filters at 77K
and at 4.2K. The tertiary mirror
is the cold Lyot stop of the instrument, sharply
limiting the field of view of the photometers.
Throat-to-throat concentrators with their entrance
aperture placed in the telescope focal plane
allow radiation to enter inside the RF-tight box 
containing the $^3$He refrigerator and the
detectors. The detectors use parabolic and conical
concentrators, metal mesh filters and Si$_3$N$_4$
spider web absorber bolometers.  
A photograph of the focal plane array for the Antarctic
flight and of its ancillary hardware is shown in fig.3.

\begin{figure}[ht!]
%\centerline{\epsfig{file=fig3.eps,height=4.5in,width=7.0in}}
\centerline{\epsfig{file=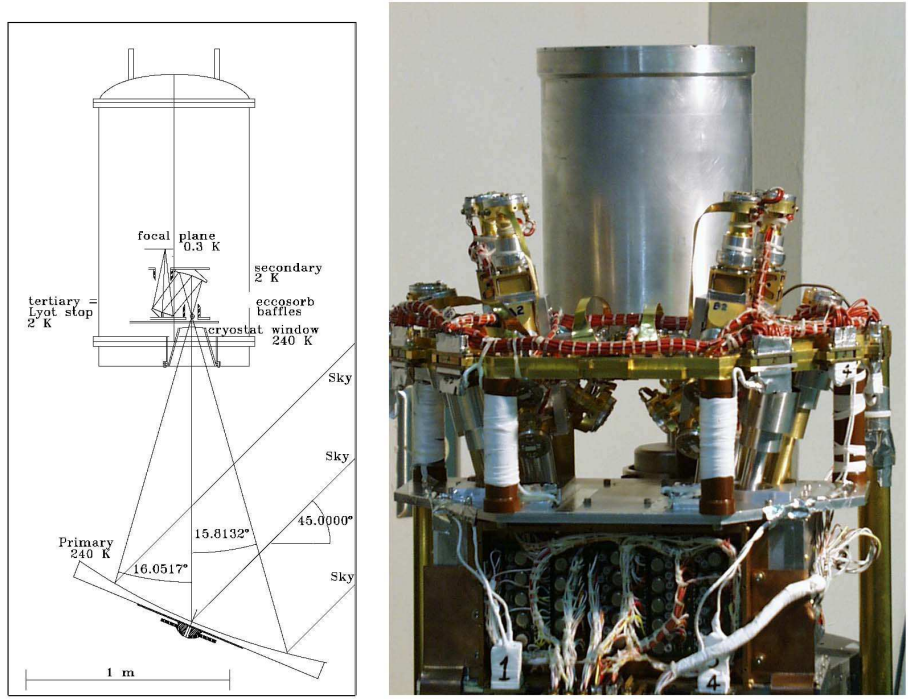}}
\caption{ Optical layout of the instrument (left) and photo of the 
focal plane array for the LDB flight. The multiband photometer
arrays are visible, mounted on the 0.3 K flange. The 0.3K stage
is mounted on four vespel columns over the cold FET stage 
visible in the lower part of the picture. The long duration
$^3$He fridge is visible behind the detectors. The height
of the cold insert shown in the picture is 50 cm.}
\label{fig3}
\end{figure}

We have two different configurations for the focal plane.
For the short test flight we traded angular resolution for throughput
in order to get significant sensitivity during the short flight from
Texas (6 hours). The final configuration, used for the Antarctic
flight, has higher angular resolution and improved detectors
(see fig.2 and 3). 
Low background bolometers are extremely sensitive to
all forms of radiant energy. One big problem with standard
bolometers is energy deposition from cosmic rays (CR) particles.
Previous experiments carried out in temperate regions
measured typical rates of an event every few seconds.
The problem is enhanced in the polar stratosphere, where we 
expect a flux of cosmic rays about 6 times larger than in the
temperate stratosphere. Bolometers with negligible
cross sections for cosmic rays have been developed \cite{bock}, \cite{Mausk}
and tested in flight. The absorber is a thin web 
($\sim 1 \mu m$ thickness)
micromachined from Si$_3$N$_4$ and metalized. 
The grid constant is of the order of a few hundred microns, 
smaller than the in band wavelength so that
photons are effectively absorbed while CR are not.
Also the heat capacity of the bolometer is greatly reduced,
and the mechanical resonant frequency is in kHz range.
NEPs of the order of $10^{-17} W/\sqrt{Hz}$ at 0.3K are achieved
with time constants of the order of 10 ms for the 150GHz
detectors. The other form of energy which contaminates
signals from high sensitivity bolometers is radio frequency
(RF) pickup, either through the optical path or 
through the readout circuit.
This problem is very severe in the case of balloon borne
payloads, which have powerful GHz telemetry transmitters on
board. The BOOMERanG receiver has been carefully shielded: all the
detectors operate in a RF tight cavity. Millimeter-wave radiation enters the 
cavity through apertures much smaller than the RF wavelength 
(inside the throat to throat parabolic concentrators), 
and all the wires enter the cavity through cryogenic EMI filter 
feedthroughs. Similar care has been taken for the signal 
processing circuits.

The bolometers are differentially AC biased and read-out.
The bolometer signals are demodulated by lock-ins synchronous
with the bias voltages. A RC cut-off at a frequency (15 mHz) lower
than the scan frequency was used to limit the dynamical range of
the data and to remove 1/f noise from very low frequencies,
unimportant for our measurements.

A long duration $^4He$ cryostat \cite{MasiB}
cooling the experimental insert to 2K, and a long
duration $^3He$ fridge \cite{MasiA}, cooling 
the photometers to 280 mK, have been developed.
The cryostat has a central 50 liter volume available
for the 2K hardware, including blocking filters,
secondary and tertiary mirrors, cryogenic preamplifiers,
back to back concentrators, $^3$He fridge, and photometers.
The cryogens hold time exceeds 12 days under flight conditions. 

The telescope and receiver hardware are mounted on a tiltable
frame (the inner frame of the payload). The observed elevation
can be selected by tilting the inner frame by means of a 
linear actuator. The outer frame of the payload is connected 
to the flight train through an azimuth pivot. The observed azimuth
can be selected by rotating the full gondola around the pivot,
by means of two torque motors. The first actuates a flywheel,
while the second torques directly against the flight train.
An oil damper fights against pendulations induced 
by stratospheric wind shear or other perturbations.
The sensors for attitude control are different for night-time
flights (like the test flight in 1997) than day-time flights
(like the Antarctic flight in 1998). For the night-time flight
in Texas we had a set of 3 vibrating gyroscopes and a flux-gate
magnetometer in the feedback loop driving the sky scan. 
A CCD camera with real-time star position measurement
 provides attitude reconstruction to 1 arcmin.
For the day-time flight in Antarctica we have a coarse and a fine
sun sensor, a differential GPS, and a set of 3 laser gyroscopes.
Again, we expect to be able to reconstruct the attitude
better than 1 arcmin. 

\section*{THE TEST FLIGHT}

The system was flown for 6 hours on Aug 30, 1997, from the
National Scientific Balloon Facility in Palestine, Texas.
All the subsystems performed well during the flight:
the He vent valve was opened at float and was closed at termination,
the Nitrogen bath was pressurized to 1000 mbar,
the $^3$He fridge temperature (290mK) drifted with 
the $^4$He temperature by less than 6mK during the flight.
Pendulations were not generated during CMB scans, at a
level greater than 0.5 arcmin, and both azimuth scans 
at 2$^o$/s and full azimuth rotations (at 2 and 3 rpm) 
of the payload were performed effectively. 
The loading on the bolometers was as expected,
and the bolometers were effectively CR immune, with
white noise ranging between 200 and 300 $\mu K \sqrt{s}$
depending on the channel. 
During the flight, the system observed
Jupiter to measure the beam pattern and responsivity. 
A strip of sky at high galactic latitudes ($\sim$ 4 deg high 
in declination, $\sim$ 6 hours wide in RA) in the constellations 
of Sagittarius, Capricornus, Aquarius, and Sculptor, was scanned 
in search for CMB anisotropies. The gondola performed
40$^o$ peak to peak azimuth scans, centered on the south, 
with 50 s period, for 5 hours. Sample scans are shown in fig.4.
The signal from the pitch gyroscope is fourier analyzed in
fig.5, showing very small pendulations of the system.
\begin{figure}[ht!]
\centerline{\epsfig{file=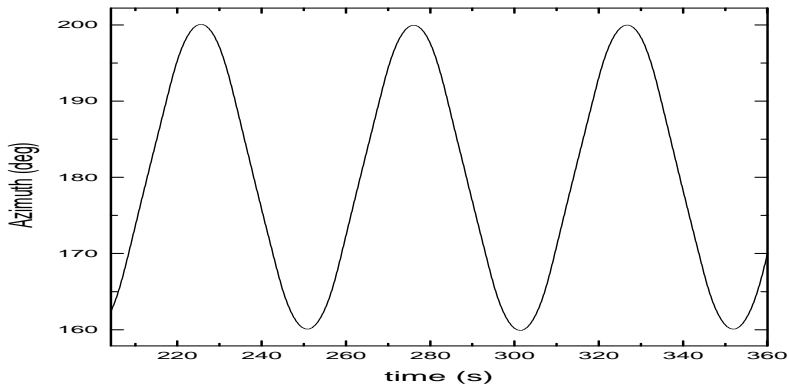,height=2.5in,width=4.5in}}
\caption{ Azimuth scans of the instrument during the test flight.
The scans are centered around south. The earth rotation
during the 5 hours of scans produced an elongated sky coverage,
4 degrees high in declination, ranging from $\sim$ 10 to 80 degrees
of Galactic latitude.}
\label{fig4}
\end{figure}
\begin{figure}[ht!]
\centerline{\epsfig{file=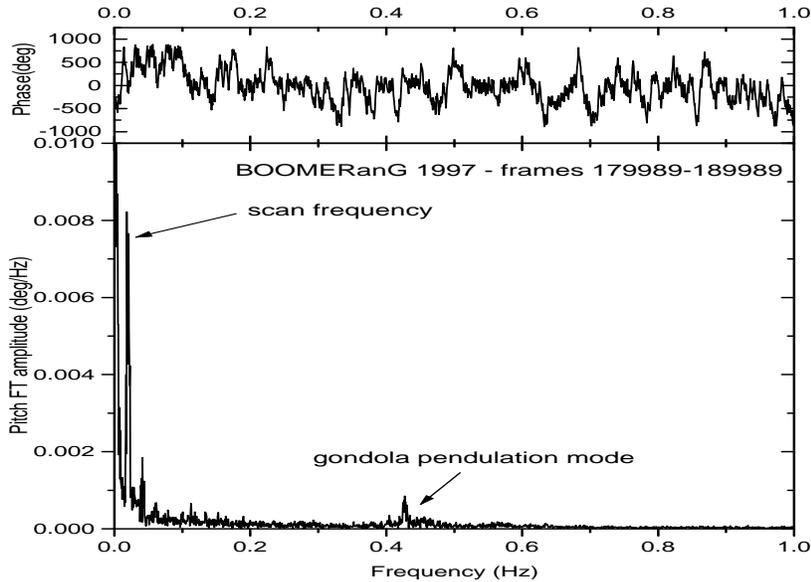,height=3.5in,width=4.5in}}
\caption{ Fourier Transform of the pitch rate gyroscope 
during the regular (40$^o$p-p) azimuth scans of the instrument.
The pendulations induced during the scans are very small.}
\label{fig5}
\end{figure}

\section*{IN-FLIGHT CALIBRATION}

The bolometers time constants turned out to be quite long
(several tens of ms) in the test flight, and we performed a 
careful analysis 
of the instrument transfer function to take care of these
effects. The electronics (AC bias readout) transfer function was 
measured accurately and found to be very stable with
temperature. The transfer function of the bolometers
depends on the reference bath temperature and on the
radiative loading, so in principle can change at float.
It has been measured in flight with two independent
methods. The first one is using cosmic rays hits. 
We have a few samples (60 hits in 6 hours in B150B2).
They are extremely reproducible, can be normalized and
synchronized. The Fourier Transform of the impulse response
gives us the total (bolometer + readout electronics)
transfer function of the instrument. To check that the
results of this method are not confused by ionization effects,
we used a second method, i.e. we analyzed the signals from
fast scans (18$^o/s$) on Jupiter. Here we have purely optical
pulses on the bolometers. Simulation shows that beam size and
beam shape effects are negligible. So we use the Fourier
transform of these pulses to get the instrument transfer
function. From these data we can recover $\tau_{bol}$, better
than 10$\%$. This is completely acceptable, since the 
calibration measurements are quite unsensitive to $\tau_{bol}$.
In fact, the instrument was calibrated using the amplitude
of the signals observed during regular ($\sim 2^o/s$) scans
on Jupiter. The calibration constant is
$$
{\mathcal{K}}=
{ \Delta V_{planet} \over T_{CMB} } 
{ \Omega \over \Omega_{planet} }
{\int E(\nu) BB(T_{CMB},\nu) {x e^x \over e^x-1} d\nu
\over 
\int E(\nu) BB(T_{eff},\nu) d\nu }
$$
with obvious meaning of the symbols. The measured
$\Delta V_{planet}$ and $\Omega$ have opposite dependance 
on $\tau_{bol}$, so ${\mathcal{K}}$ changes by less than
1 $\%$ for a 20 $\%$ change of $\tau_{bol}$. 
We also checked with simulations that our deconvolution
from the instrument transfer function does not affect 
significantly the shape and value of CMB anisotropy signals 
in our detectors (see \cite{aquila1}). 
In fig.6 we plot an example of signal from scans on Jupiter.
\begin{figure}[htb]
\centerline{\epsfig{file=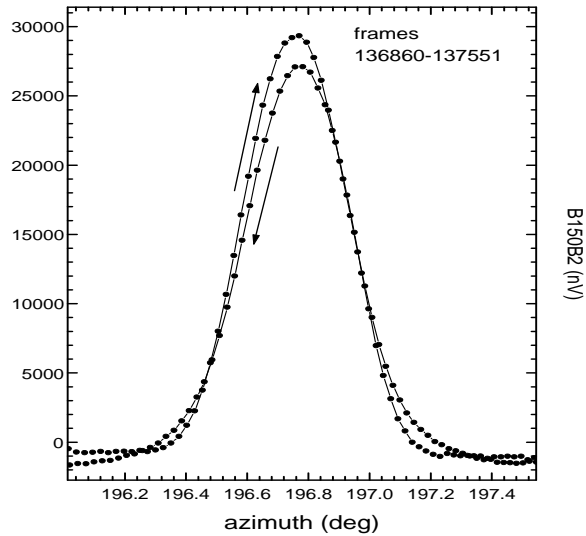,height=3.5in,width=3.5in}}
\caption{ Sample real time signals 
during azimuth scans on Jupiter. The scan speed is 1.7$^o/s$
towards increasing azimuth,
the elevation of the beam is 39$^o$. Due to the non-negligible time
constant of the bolometers, the signals are non-symmetric.
The effect of the high-pass filter is also evident in the
baseline. These effects are removed by deconvolution from 
system transfer function in the data analysis. The forward
and reverse scans shown here are different due to the
sky rotation and different payload attitude.}
\label{fig6}
\end{figure}
The responsivity calibration has been obtained with the 
following sequence of steps: signals are deconvolved from
the electronics readout and bolometer time constant, and
then is filtered using a flat phase numeric filter to 
reduce white noise and cut drifts. The best parameters 
for the filter are obtained with numerical simulations.
The filtered data are binned on a grid centered on the
optical position of Jupiter. The beam solid angle
$\Omega$ is computed by 2-D integration of the binned data. 
We get FWHM of 19.5, 19.5, 16.5, 24, 29 arcmin for the
B150A1, B150B1, B150B2, B90A, B90B channels respectively.
$\Delta V_{Jupiter}$ is estimated making fits to the scan data.
${\mathcal{K}}$ is derived using the measured quantities
and the spectral information $E(\nu)$ as measured in the
laboratory. Errors are propagated. The final precision of
the calibration ${\mathcal{K}}$ is 5.9, 8.7, 3.8,
7.0, 5.2 $\%$ for the B150A1, B150B1, B150B2, B90A, B90B
channels respectively. 

An independent method of calibration for ${\mathcal{K}}$
is the observation of the CMB Dipole. In fact, the Dipole
has the same spectrum as smaller scale CMB anisotropies,
fills the beam, and its amplitude and direction
are well known from COBE-DMR (better than 1$\%$). The Dipole
was observed during the test flight for four times, to check
unambiguously the celestial nature of the signal. On our
scans, the expected Dipole signal ranged between 2 and 4
mK$_{CMB}$ peak to peak, depending on the scan. We do see
a Dipole signal in all the 90 and 150 GHz channels. The
signal is not correlated to the (small) pendulations
sensed by the pitch and roll gyroscopes, and rotates
with the sky, thus excluding any atmospheric origin.
In fig.7 we plot sample observed signals 
during one of the azimuth spins of the payload.

\begin{figure}[ht!]
\centerline{\epsfig{file=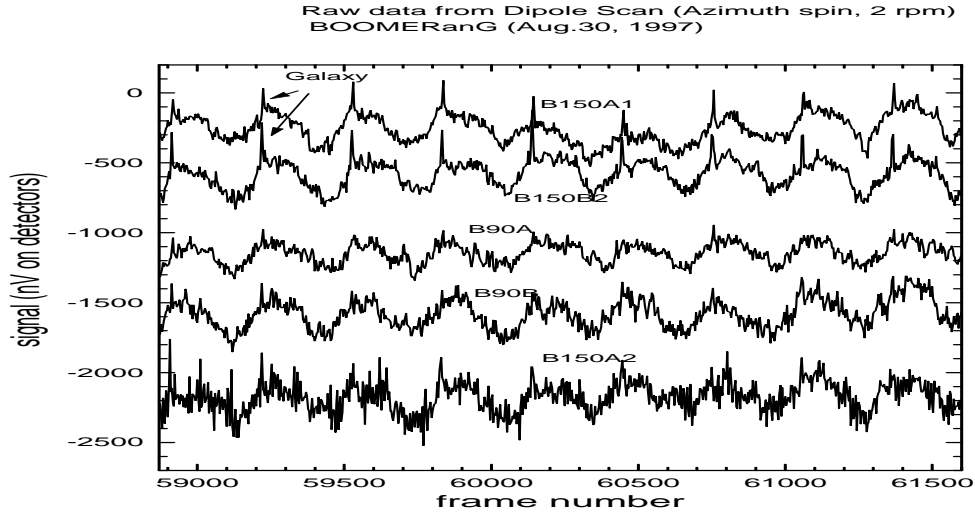,height=3.0in,width=5.5in}}
%\caption{ 90 GHz data binned in angle from the DMR CMB dipole
%direction, are plotted versus the theoretical Dipole signal
%(rotation 2). }
\caption{ Data from the active detectors during the
3 rpm rotations of the payload. Nine rotations of the
payload are shown. The CMB dipole signal is
evident in all the channels.}
\label{fig7}
\end{figure}
\begin{figure}[ht!]
\centerline{\epsfig{file=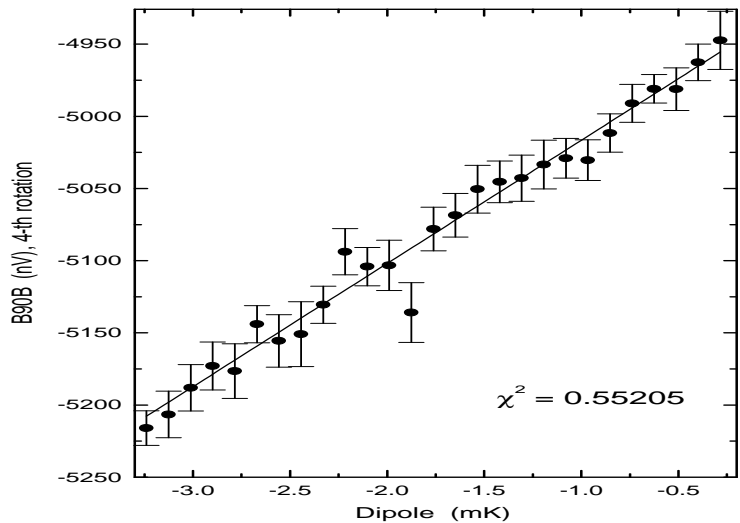,height=3.5in,width=4.5in}}
\caption{ 90 GHz data binned in angle from the DMR CMB dipole
direction, are plotted versus the theoretical Dipole signal
(rotation 2). }
\label{fig8}
\end{figure}
  
In fig.8 we plot the observed signals averaged in dipole
angle bins versus the COBE-DMR dipole for channel B90B.
The calibration ${\mathcal{K}}_D$ (derived as the ratio
between the observed dipole and the COBE-DMR dipole on the 
same path) agrees well with the ${\mathcal{K}}$ measured
from Jupiter at the same time. 
The final precision of the calibration for the 
first rotation is: $2, 4, 6, 4, 3\%$ for the B90A, B90B,
B150A1, B150B1, B150B2 channels respectively.
Calibration drifts are
evident from both the Dipole signals and from the internal
calibrator lamp, which flashed every 15 minutes and can
be used as a calibration transfer during the entire flight.
In summary, we estimate that our total calibration error is
significantly less than 10$\%$ for all the active channels.

\section*{CMB DATA ANALYSIS PIPELINE}

The $\sim 1.5 \times 10^6$ samples are first edited 
removing cosmic rays hits,
calibration lamp flashes, microphonic events from the
nitrogen pressurization system, radar hits, signals from
Jupiter, data taken in rotation mode. This editing
removes less than 5$\%$ of the data. The resulting
data set is mainly noise, and is stationary: the rms
changes less than 2$\%$ from hour to hour. Then the
data are filtered in the time domain: they are deconvolved
and flat phase filtered as described in the previous paragraph;
a notch filter at 6.8 Hz is applied to remove temperature
control bias aliases where needed. The following steps
are part of a loop which is repeated under different
assumptions (forward or reverse scans, flight sections
with different environment, etc.) to check for systematic 
effects. The noise matrix is estimated from the power
spectrum of the data. The CMB analysis continues along
two parallel paths. Along the first one, data are binned
using synthesized beam patterns (as in \cite{Net}), and
the corresponding window functions and the covariance
 matrix of the data are computed. A likelihood analysis
produces band-power estimates of the $c_\ell$. Along the
second path an "optimal" map is created from the data stream
following the procedures described in \cite{map}, and the
$c_\ell$ power spectrum is computed from the map
(see e.g. the papers from Borrill and Jaffe in these proceedings).
All this is repeated with different data selection rules, 
deconvolution parameters, noise estimate etc., to check  
for systematic effects. The procedure is applied to all
the channels and to the dark channel, and correlation analysis
are carried out to check for other systematic effects.
It is worth to stress the following facts:  
the BOOMERanG 1997 data set produces
a 26000 pixels map (6 arcmin pixelization with a 15 arcmin
FWHM beam). At the moment, this is the biggest CMB map ever made.
At the pixel level, the map is dominated by noise. The map
contains both dusty sections (closer to the Galactic plane)
and clean sections (up to $b \sim 80^o$). Both are
interesting. Power spectra have been made, and we do have
significant detections of CMB anisotropies in the range 
$50 \simlt \ell \simlt 400$. The data are sample variance
limited in this range.
We are in the middle of the "check for systematics loop" 
process, so we cannot give final results yet. However,
the two methods outlined give consistent results for
the measured $c_\ell$s. 

\section*{THE LDB FLIGHT}

BOOMERanG-LDB has been qualified for flight on Aug. 25, 1997.
A photograph of the payload during the qualification is visible
in fig.9.
\begin{figure}[ht!]
%\centerline{\epsfig{file=fig9.eps,height=7.0in,width=4.5in}}
\centerline{\epsfig{file=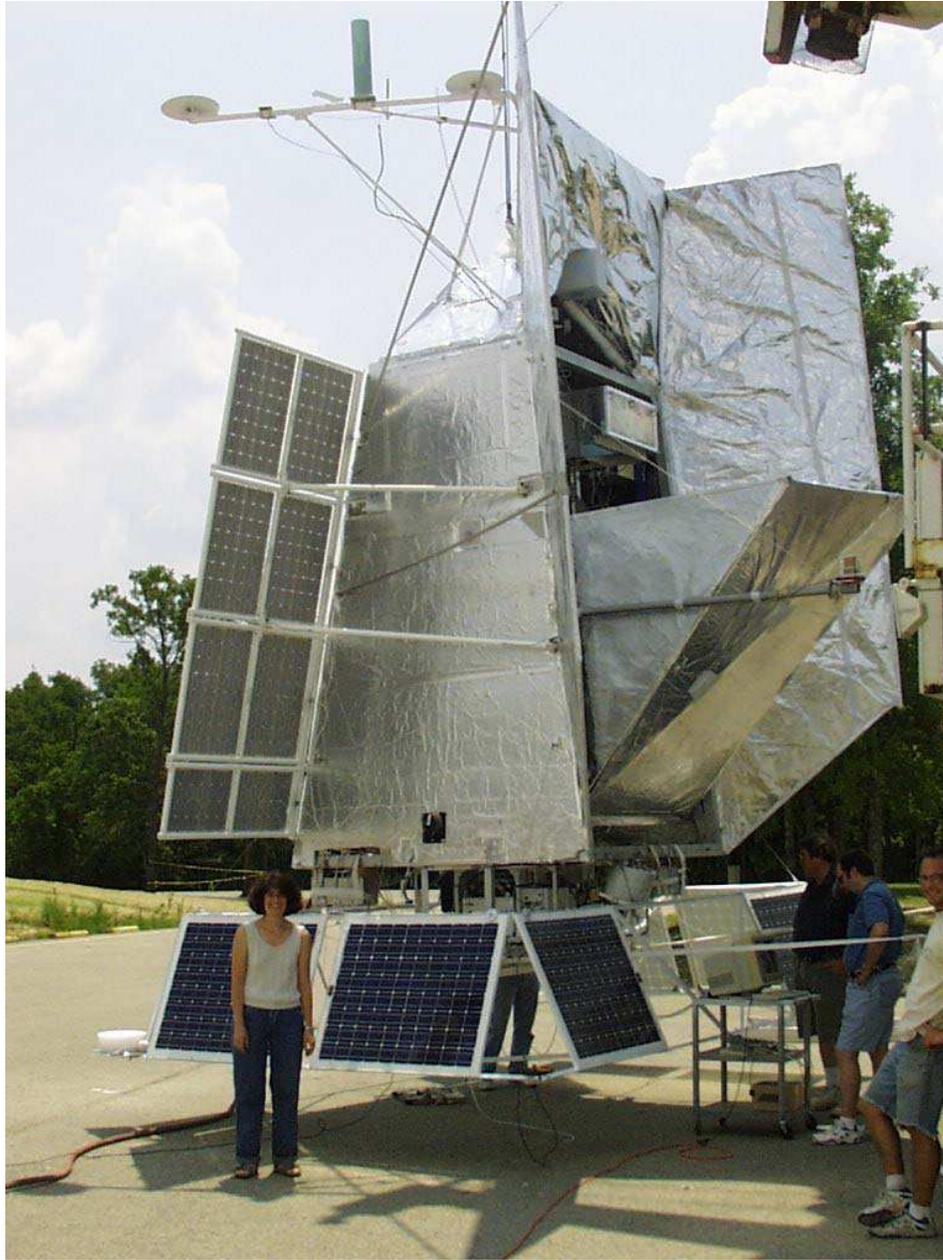}}
\caption{ The BOOMERanG LDB payload during the qualification
test at the NASA-NSBF in Palestine (Texas) on August 25, 1997.
The earth shield and the sun shields dominate the scene.
Also the solar array providing the power supply for the
payload (upper) and for the navigation hardware
(lower) are evident. The payload is 5.2 m high and the total
weight is $\sim 1.5$ ton.}
\label{fig9}
\end{figure}
The main differences with respect to the BOOMERanG-NA,
are the new focal plane and the improved detectors.
Now the measured time constant for the detectors range
between 16 ms (for the 90 GHz channels) and 2 ms (for the
400GHz channels). Coarse and Fine Sun sensors have been
developed and calibrated for pointing reconstruction.
The gyroscopes driving the scan have been replaced with
optical fiber ones, featuring very low drift.
The main sensors for pointing reconstruction, resetting
the rate gyroscopes, are the fine sun sensors, two digital
sundial featuring subarcminute resolution with a measurement
range of $\simgt 120^o$ in azimuth and 50$^o$ in elevation.
A differential GPS completes the sensors set.
An array of solar panels, providing $\sim 2 kW$ of
maximum power, has been implemented on the payload.
Real time monitoring of most of the instrument signals
will be possible during all the LDB flight using the
TDRSS satellites constellation.
The instrument will be calibrated against several HII
regions visible in the southern Hemisphere, close to
the main target of the instrument, the Horologium
constellation, and far from the Sun. The same regions
will be mapped from ground using the SUZIE instrument.

\section*{CONCLUSIONS}

The test flight of the BOOMERanG experiment has demonstrated
the good performance of the instrument and has qualified
it for the LDB flight. The instrument has been precisely
calibrated and CMB anisotropies have been detected in the
range of the first acoustic peak of the $c_\ell$ spectrum.
The data analysis pipeline has been developed, and
we are currently performing deep checks for systematic 
effects. The LDB instrument has been shipped to Antarctica
for flight at the end of 1998.

\section*{Acknowledgements}

This paper merges the presentations given by S. Masi and 
P. de Bernardis. The BOOMERanG experiment has been funded by 
Programma Nazionale di Ricerche in Antartide (PNRA),
Universita' di Roma La Sapienza, Agenzia Spaziale Italiana (ASI) 
in Italy, NASA and Center for Particle Astrophysics in USA, 
PPARC in UK. We are very gratefull to the National Scientific 
Balloon Facility in Palestine, Texas for the 1997 test flight
and for professional and effective support during qualification
for the LDB. 

{\bf Note added in proof} The LDB payload has been
successfully launched by NSBF on Dec.29, 1998, and flown until 
Jan.8, 1999, from the base of William Field (Antarctica).
The system performed nominally, with detector noise on the order
of 100 $\mu K \sqrt{s}$. We accumulated 259 hours of very good 
quality data at float. The payload has been recovered in good shape.


\begin{references}

\bibitem{delab} Delabrouille J., Gorski K.M., Hivon E., {\it Mon.\ Not.\ Roy.\ Astron.\
Soc.} {\bf 298}, 445 (1998). 

\bibitem{Lange1} Lange  et al. Space Science Rev.,{\bf 74}, 1-2, (1995)

\bibitem{Paolo1} de Bernardis et al., proceedings of the meeting 
Microwave Background Anisotropies, Moriond Astrophysics Meeting, 
F. Bouchet editor, Ed. Frontieres, 1996, pg. 155.

\bibitem{Silvia1}  Masi et al.  Proceedings of the workshop 
Topological Defects in Cosmology, Rome, 1996, 
M. Signore and F. Melchiorri editors (World Scientific)

\bibitem{Schl} D.J. Schlegel, D.P. Finkbeiner, M. Davis, ApJ, 500, 525, 20

\bibitem{bock} Bock J., Chen P., Mauskopf P., Lange A., \emph{A novel bolometer
for infrared and millimeter-wave astrophysics}, Space Sci. Rev.,
{\bf 74}, 229-235, 1995

\bibitem{Mausk} Mauskopf P., et al., \emph{Composite infrared bolometers with $Si_3N_4$
micromesh absorbers}, Applied Optics, {\bf 36}, 765-771, 1997. 

\bibitem{MasiA} S. Masi, E. Aquilini, P. Cardoni, 
P. de Bernardis, L. Martinis, F. Scaramuzzi, D. Sforna, 
Cryogenics, 38, 319-324, 1998A

\bibitem{MasiB} S. Masi, P. Cardoni, P. de Bernardis, 
F. Piacentini, A. Raccanelli, F. Scaramuzzi, 
"A long duration cryostat suitable for balloon borne photometry"
1999, Cryogenics, in press

\bibitem{aquila1} P. de Bernardis, G. De Troia, L. Miglio, 1999,
Proc. of the International School of Space Science: coures
"3K Cosmology from Space", F. Melchiorri, M. Signore, P. Richards, 
G. Sironi editors, Elsevier.

\bibitem{Net} Netterfield C.B. et al., Ap.J., {\bf 474}, 47, 1997 

\bibitem{map} see e.g. Janssen, astro-ph 9602009, 
Wright astro-ph 9612006, Smoot astro-ph 9704193, 
Tegmark astro-ph 9711076. 

\end{references}
\end{document}